\documentclass[lineno]{jfm}

\usepackage{graphicx}
\usepackage{amsmath}
\usepackage{amssymb}
\usepackage{pifont}
\usepackage{txfonts}
\usepackage{psfrag}
\usepackage[round]{natbib}
\usepackage[T1]{fontenc}
\usepackage{lmodern}
\usepackage{tikz}
\usetikzlibrary{shapes.geometric,shapes.symbols}
\usepackage{scalerel}
\usepackage{upgreek}
\usepackage{booktabs}
\usepackage{verbatim}
\usepackage{framed}
\usepackage{fancyvrb}
\usepackage{varwidth}
\usepackage{bbding}
\usepackage{subfigure}
\usepackage{multirow}
\usepackage[normalem]{ulem}
\usepackage{multicol}
\usepackage{xcolor}
\usepackage{breqn}
\usepackage{soul}
\usepackage{dirtytalk}
\usepackage{upquote}
\usepackage{authblk}
\usepackage{array}
\usepackage{lscape}
\usepackage{pdflscape}

\newcolumntype{P}[1]{>{\centering\arraybackslash}p{#1}}
\definecolor{violet}{rgb}{0.54, 0.17, 0.89}
\definecolor{golden}{rgb}{0.99, 0.76, 0.0}

\usepackage{hyperref}
\hypersetup{
    colorlinks = true,
    urlcolor   = blue,
    citecolor  = black,
}

\newcommand{\RomanNumeralCaps}[1]
\linenumbers

\newcommand{\rahul}[1]{{\color{black}#1}}

\title{ {Active and inactive contributions to the wall pressure and wall-shear stress in turbulent boundary layers}}

\author{Rahul Deshpande\aff{1}
  \corresp{\email{raadeshpande@gmail.com}},
Ricardo Vinuesa\aff{2},
Joseph Klewicki\aff{1}
and Ivan Marusic\aff{1}}

\affiliation{\aff{1}Dept. Mechanical Eng., University of Melbourne, Parkville, VIC 3010, Australia
\aff{2}FLOW, Eng. Mechanics, KTH Royal Institute of Technology, Stockholm, 10044, Sweden}

\begin{document}
\maketitle

\begin{abstract}

A phenomenological description is presented to explain the intermediate and low-frequency/large-scale contributions to the wall-shear-stress (${\tau}_w$) and wall-pressure (${p}_w$) spectra of canonical turbulent boundary layers, which are well known to increase with Reynolds number.
The explanation is based on the concept of active and inactive motions 
(\citeauthor{townsend1961}, \textit{J.\ Fluid Mech.}, vol.\ 11, 1961) associated with the attached-eddy hypothesis.
Unique data sets of simultaneously acquired ${\tau}_w$, ${p}_w$ and velocity fluctuation time series in the log region are considered, across friction-Reynolds-number ($Re_{\tau}$) range of $\mathcal{O}$($10^3$) $\lesssim$ $Re_{\tau}$ $\lesssim$ $\mathcal{O}$($10^6$). 
A recently proposed energy-decomposition methodology (\citeauthor{deshpande2021}, \textit{J.\ Fluid Mech.}, vol.\ 914, 2021) is implemented to reveal the active and inactive contributions to the ${\tau}_w$- and $p_w$-spectra. 
Empirical evidence is provided in support of \citeauthor{bradshaw1967}'s (\textit{J.\ Fluid Mech.}, vol.\ 30, 1967) hypothesis that the inactive motions are responsible for the non-local wall-ward transport of the large-scale inertia-dominated energy, which is produced in the log region by active motions.
This explains the large-scale signatures in the ${\tau}_w$-spectrum, \rahul{which grow with $Re_{\tau}$ despite the statistically weak signature of} large-scale turbulence production, \rahul{in the near-wall region}.
For wall pressure, active and inactive motions respectively contribute to the intermediate and large scales of the $p_w$-spectrum. 
Both these contributions are found to increase with increasing $Re_{\tau}$ owing to the broadening and energization of the wall-scaled (attached) eddy hierarchy.
\rahul{This potentially explains the rapid $Re_{\tau}$-growth of the $p_w$-spectra relative to ${\tau}_w$, given the dependence of the latter only on the inactive contributions.}

\end{abstract}

\begin{keywords}
turbulent boundary layers, boundary layer structure.
\end{keywords}

\section{Introduction and motivation}
\label{intro}

Wall-bounded flows are ubiquitous in various applications, such as over aircraft wings and around submarines, where they affect vessel performance through the imposition of wall-shear-stress (${\tau}_{w}$) and wall-pressure ($p_{w}$) fluctuations on the bounding surface.
The former, ${\tau}_w$ are associated with the skin-friction drag which limits vehicle speed, while the latter, $p_w$ are responsible for flow-induced vibrations that affect structural stability and generate noise.
 {Nevertheless, accurately modelling the fluctuations in ${\tau}_w$ and $p_w$ remains 
a challenge due to the limited understanding of dependence of their generating mechanisms on the Reynolds number.}

\begin{figure}
   \captionsetup{width=1.0\linewidth}
\centering
\includegraphics[width=1.0\textwidth]{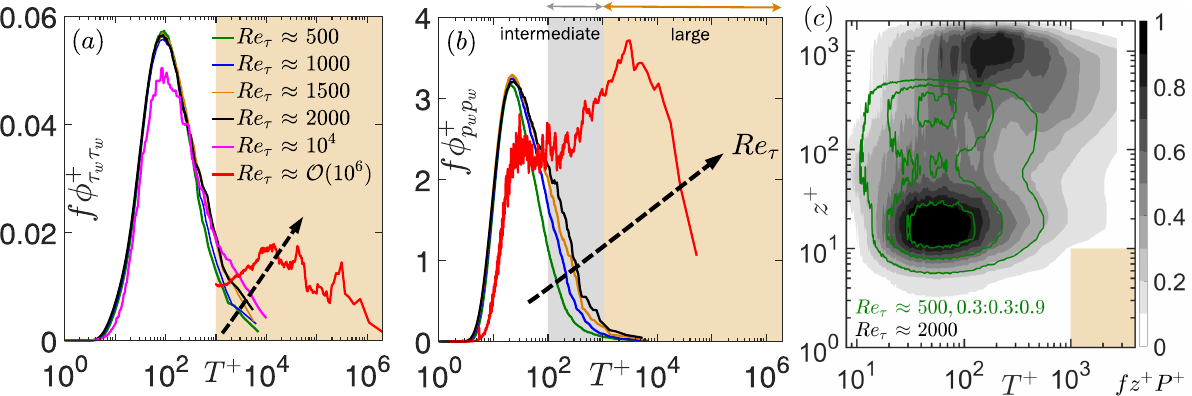}
\caption{Premultiplied energy spectra of (a) ${\tau}_w$, (b) $p_w$ and (c) premultiplied spectrogram of the bulk turbulence production ($f{z^+}{P^+}$) as a function of $T^+$ (= $1/{f^+}$).
Data for 500 $\lesssim$ $Re_{\tau}$ $\lesssim$ 2000 are from \citet{eitel2014}, at $Re_{\tau}$ $\sim$ $\mathcal{O}$($10^4$) is from \citet{marusic2021}, and that at $Re_{\tau}$ $\sim$ $\mathcal{O}$($10^6$) is from the SLTEST data acquired by (a) \citet{marusic2007} and (b) \citet{klewicki2008}.
Grey and yellow coloured backgrounds in (a-c) indicate the intermediate ($\mathcal{O}$($10^2$) $\lesssim$ $T^+$ $\lesssim$ $\mathcal{O}$($10^3$)) and large scale ranges ($T^+$ $\gtrsim$ $\mathcal{O}$($10^3$)), respectively.}
\label{fig1}
\end{figure}

Figure \ref{fig1} shows a compilation of premultiplied frequency spectra of ${\tau}_{w}$ and $p_{w}$
for a wide range of Reynolds numbers. 
Here, the viscous-scaled time is defined as $T^+$ = ${{{U}^2_{\tau}}}/({f}{\nu})$, where $f$ is the frequency of turbulence scales, $\nu$ is the kinematic viscosity and ${U}_{\tau}$ is the mean friction velocity, with superscript `+' indicating viscous scaling.
All these spectra have been computed from previously published high-fidelity simulations (500 $\lesssim$ $Re_{\tau}$ $\lesssim$ 2000) and experimental data sets ($\mathcal{O}$($10^4$) $\lesssim$ $Re_{\tau}$ $\lesssim$ $\mathcal{O}$($10^6$)) of a turbulent boundary layer (TBL), where $Re_{\tau}$ = ${U}_{\tau}{\delta}/{\nu}$ and $\delta$ is the TBL thickness.
 {Although the experimental spectra have certain limitations at $T^+$ $\lesssim$ 100 owing to spatial and temporal resolution, they will not influence this discussion} (see $\S$\ref{data}).

A noteworthy observation from figures \ref{fig1}(a,b) is the increasing energy contribution to the ${\tau}_w$- ($T^+$ $\gtrsim$ $\mathcal{O}$($10^3$)) and $p_w$-spectra ($T^+$ $>$ $\mathcal{O}$($10^2$)) with increasing $Re_{\tau}$, which has been noted previously in all canonical wall flows \citep{tsuji2007,orlu2011,mathis2013,panton2017,yu2022,baars2023}.
While past studies have linked the trends in $Re_{\tau}$ to the energization of large inertial motions, this study addresses two fundamental questions that have remained unanswered:
(i) How does the energy of the large-scales (hereafter, large-scale energy), residing predominantly in the outer region \citep{mklee2019}, propagate towards the wall to influence ${\tau}_w$?
(ii) Which motions contribute to the rapid $Re_{\tau}$ growth of $p_w$ in the intermediate scales ($\mathcal{O}$($10^2$) $\lesssim$ $T^+$ $\lesssim$ $\mathcal{O}$($10^3$)), \rahul{which is however insignificant for the case of} ${\tau}_w$?

The origin of the large-scale inertia-dominated energy in the outer region has now been well established in the literature \citep{mklee2019}, based on the turbulent kinetic energy (TKE) production term, $P^+$ = -2${{\overline{uw}}^+}({{\partial}{{{U}}^+}}/{{\partial}{z^+}})$.
Here, $u$, $v$ and $w$ represent instantaneous velocity fluctuations along the streamwise ($x$), spanwise ($y$) and wall-normal ($z$) directions respectively, while capital letters and overbar indicate time-averaging.
Figure \ref{fig1}(c) shows the premultiplied spectra of the bulk turbulence production (\rahul{$f{z^+}{P^+}$($z^+$; $T^+$)}) from the same simulation data sets as in figures \ref{fig1}(a,b).
\rahul{Note here that the premultiplication with $z^+$ artificially amplifies the  spectral content in the outer region compared to the inner region, but the majority of the energy production essentially occurs near the wall when $Re_{\tau}$ $\lesssim$ $\mathcal{O}$($10^3$).
Figure \ref{fig1}(c) confirms the weak energy production in the near-wall large-scales ($T^+$ $>$ 1000) compared to the small-scales, thereby suggesting association of the large-scale ${\tau_w}-$signatures with wall-ward energy transport from the outer region \citep{mklee2024}.}
While several past studies have quantified and predicted the superposition of large-scale signatures on ${\tau}_w$ \citep{metzger2001,orlu2011,mathis2013}, fundamental understanding of their energy-transfer mechanisms, particularly the transport  of large-scale energy from outer region towards wall, is still lacking.

With regards to the differences between ${\tau}_w$ and $p_w$ over $\mathcal{O}$($10^2$) $\lesssim$ $T^+$ $\lesssim$ $\mathcal{O}$($10^3$), although considerable TKE is produced in this scale range in the inner region, \rahul{the contour levels of the TKE production spectra in figure \ref{fig1}(c) exhibit a convincing collapse/invariance across different $Re_{\tau}$. 
This suggests a negligible increment in the near-wall TKE production for increasing $Re_{\tau}$, consistent with \citet{mklee2019}}. 
It means that the $Re_{\tau}$-increase of $f{{\phi}^{+}_{{{p}_w}{{p}_w}}}$ in the intermediate-scale range is also \rahul{associated with} the energization of turbulence motions farther away from the wall, which, however, \rahul{seem to influence $f{{\phi}^{+}_{{{\tau}_w}{{\tau}_w}}}$ very weakly.
The present study attempts to advance our fundamental understanding of the turbulence motions governing these trends, and their roles in the underlying energy-transfer mechanisms.}
This information would be valuable for designing high-$Re_{\tau}$ drag and noise reduction strategies in the future. 


\section{ \rahul{Active and inactive components and their roles in the energy-transfer mechanisms}}
\label{aem}

\begin{figure}
   \captionsetup{width=1.0\linewidth}
\centering
\includegraphics[width=1.0\textwidth]{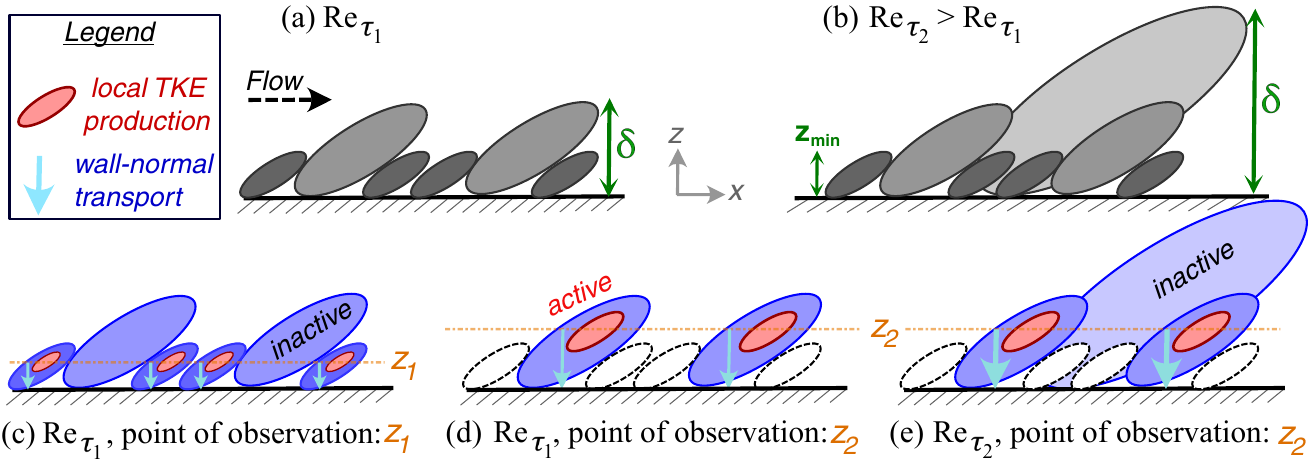}
\caption{ {Conceptual sketches of wall-scaled eddies in a TBL at relatively (a) low ($Re_{{\tau}_1}$) and (b) high ($Re_{{\tau}_2}$) Reynolds numbers. 
(c-e) depict the roles of the active and inactive parts of wall-scaled eddies in the energy-transfer mechanisms per hypothesis of \citet{bradshaw1967} and \citet{giovanetti2016}. 
Blue and red sections respectively correspond to the inactive and active portions of the wall-scaled eddies (grey-scaled), which are defined based on the point of observation, $z$: (c) $z_1$ or (d,e) $z_2$.
In (d,e), white eddies with dashed contours are centred lower than $z$, and hence are not involved in either local TKE production at $z$ (red region), or non-local energy transport from $z$ to the wall (cyan arrows).
Greater thickness of arrows in (e) indicates larger transport magnitude at higher $Re_{\tau}$.}}
\label{fig2}
\end{figure}

This study explores the ability of the attached-eddy (\emph{i.e.}, wall-scaled eddy; \citealp{townsend1976}) framework to provide a phenomenological explanation for the questions raised above. 
This is motivated by the fact that the same framework has been previously successful in predicting the ${\tau}_w$- and $p_w$-spectra, associated with the inertial scales, across a broad $Re_{\tau}$-range \citep{marusic2021,ahn2010}.
Per Townsend's hypothesis, the outer region ($z$ $\gtrsim$ $z_{\rm min}$) of a high-$Re_{\tau}$ wall-bounded flow can be statistically represented by a hierarchy ($\mathcal{O}$($z_{\rm min}$) $\lesssim$ $\mathcal{H}$ $\lesssim$ $\mathcal{O}$($\delta$)) of geometrically self-similar, inviscid, wall-scaled eddies dominated by inertia (figure \ref{fig2}), where $\mathcal{H}$ is the eddy height and \rahul{$z_{\rm min}$ is the nominal lower bound of the log region.}
These wall-scaled eddies have been classically referred to as attached eddies in the literature, in reference to their scaling with $z$ and ${U}_{\tau}$, and not necessarily implying their physical extension to the wall.
\citet{townsend1976} proposed that these eddies have a population density varying inversely to their height $\mathcal{H}$, leading to their cumulative velocity contributions at $z$ $\gtrsim$ $z_{\rm min}$ to follow (at asymptotically high $Re_\tau$):
\begin{equation}
\begin{aligned}
\label{eq1}
{{\overline{u^2}}^+} &= {B_1} - {A_1}\; \ln({z}/{\delta}), \\
{{\overline{v^2}}^+} &= {B_2} - {A_2}\; \ln({z}/{\delta}),  \\
{{\overline{w^2}}^+} &= {B_3}\; \text{and} \\
{{\overline{uw}}^+} &= {B_4},
\end{aligned}
\end{equation}
where $B_{1-4}$ and $A_{1-2}$ are constants.
The increased availability of high-$Re_{\tau}$ data over the past two decades have provided considerable empirical support to these expressions \rahul{\citep{jimenez2008,hultmark2012,orlandi2015,mklee2015} in the nominal log-region of the TBL (the exact definition of which may vary in the literature)}.
The disagreement beyond the log region is owing to the statistical significance of the non-wall-scaled eddy-type motions coexisting in boundary layers, particularly in its wake region \citep{marusic1995}.

While the expressions for ${\overline{u^2}}^+$ and ${\overline{v^2}}^+$ in (\ref{eq1}) suggest their dependence on $Re_{\tau}$, ${\overline{w^2}}^+$ and ${\overline{uw}}^+$ are postulated to be $Re_{\tau}$-independent \rahul{in the large $Re_{\tau}$ limit}.
\citet{townsend1976} explained this contradiction by proposing that the flow at any point of observation ($z$ $\gtrsim$ $z_{\rm min}$)  {comprises of two components of wall-scaled eddies} (figure \ref{fig2}) $-$ an `active' component  {of wall-scaled eddy} that is responsible for \emph{local} turbulent transfer and accounts for the instantaneous Reynolds shear stresses ($uw$) at $z$, and an `inactive' component that does not contribute to $uw$ at $z$.
 {Here, the inactive component predominantly comes from wall-scaled eddies} much taller than $z$, which contribute to $uw$ (\emph{i.e.}, behave as active motions) at wall-normal locations larger than $z$.
Hence, while the active component is solely responsible for the \emph{local} TKE production, \citet{townsend1976} described inactive motions as \emph{non-local} `swirling motions' whose ``\emph{effect on that
part of the layer between the point of observation and the wall is one of slow random
variation of `mean velocity’ which cause corresponding variation of wall stress.}''
 {These statements were supported by \citet{bradshaw1967} and \citet{giovanetti2016}, who hypothesized that the inactive parts of the wall-scaled eddies are responsible for transporting large-scale energy produced in the outer layer (by active parts), to the wall.}
As per \citet{bradshaw1967}, this process is necessary to maintain the near-wall energy balance and is sketched in figures \ref{fig2}(c-e).

\rahul{Although the non-local wall-normal transport of large-scale energy has been previously observed through spectral analysis of the TKE-budget equations \citep{cho2018,mklee2019}, its connection with the inactive part of the wall-scaled eddies has never been definitively established, primarily due to the lack of a reliable flow-decomposition methodology.
Along the same lines, while the association of $p_w$-signatures with intense $uw$ events is well known \citep{gibeau2021,baars2023},  {their correlation with $uw$-contributing (active) and non-contributing (inactive) components has never been explicitly established.}
Recently, \citet{deshpande2021} proposed a spectral linear stochastic estimation (SLSE)-based methodology that enables data-driven decomposition of the log-region flow into their corresponding active and inactive components.
This provides an opportunity to establish the association of active/inactive components with ${\tau}_w$ and $p_w$, and thereby explain their $Re_{\tau}$-variation based on the $Re_{\tau}$-growth of the  {wall-scaled eddy} hierarchy (figure \ref{fig2}).}

\begin{figure}
   \captionsetup{width=1.0\linewidth}
\centering
\includegraphics[width=1.0\textwidth]{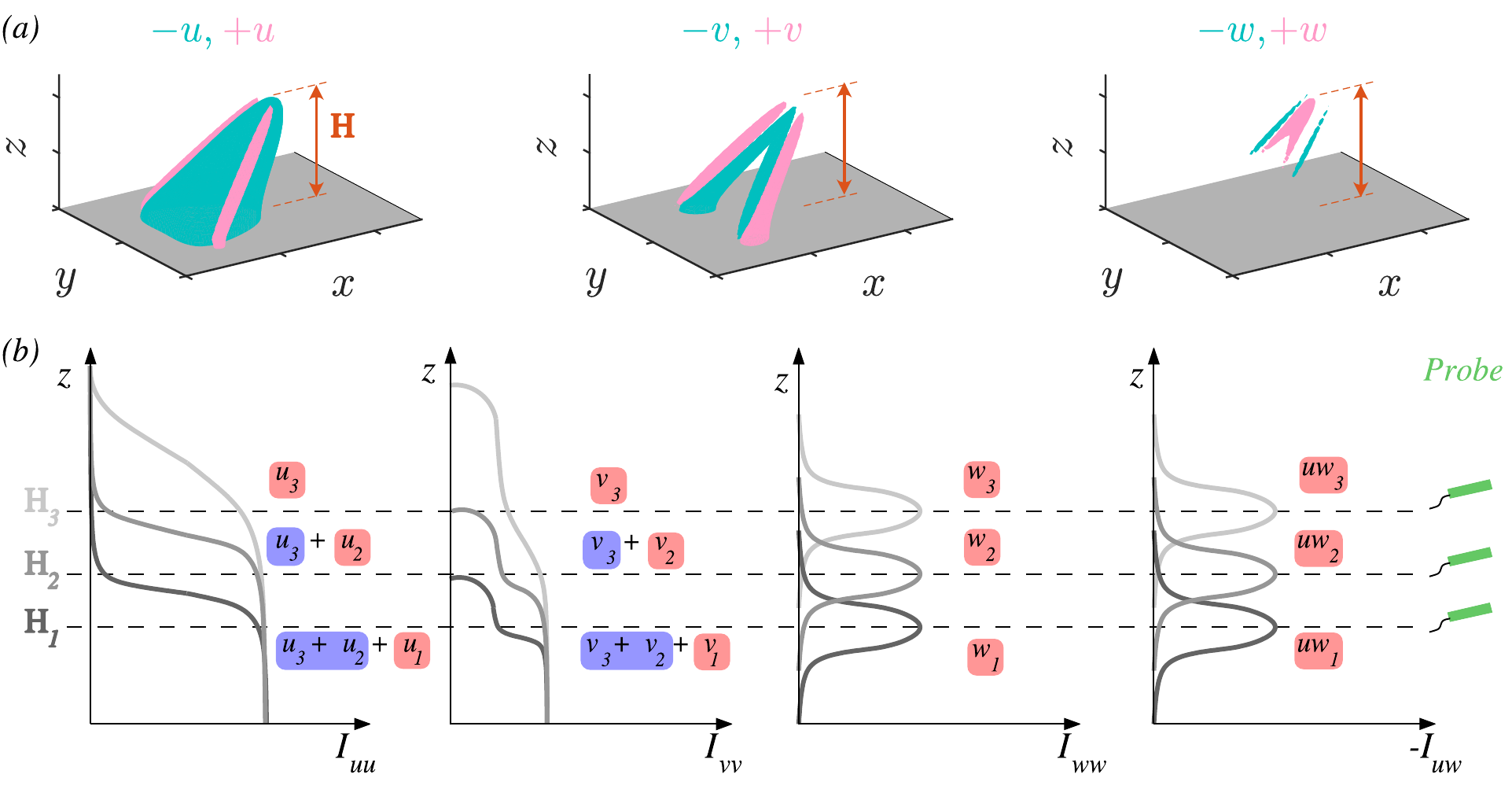}
\caption{\rahul{(a) Spatial signatures of the three velocity components ($u$,$v$,$w$) from a hairpin-type vortex structure of height $\mathcal{H}$, representative of a wall-scaled eddy \citep{marusic1995}. 
Regions in magenta and indigo respectively denote high- and low-momentum regions for the corresponding velocity fluctuations.
(b) Eddy-intensity functions ($I_{ij}$) for wall-scaled eddies of three different heights ${\mathcal{H}}_i$, with ${\mathcal{H}}_1$ $<$ ${\mathcal{H}}_2$ $<$ ${\mathcal{H}}_3$.
$u_i$,$v_i$,$w_i$ respectively denote velocity signatures generated from these wall-scaled eddies that are sensed by a probe, depending on their $z$-location.
Red and blue background shading of $u_i$,$v_i$,$w_i$ respectively indicates  active and inactive contributions at that location.
The figure has been adapted from \citet{deshpande2020afmc}.}}
\label{fig3}
\end{figure}

\subsection{ \rahul{Revisiting the concept of active and inactive motions}}
\label{concept}

\rahul{Before proceeding with the new analysis, we briefly review the concept of active and inactive motions by means of a simplified mathematical description, which is inspired by Townsend's original ideas and adapted from our previous work \citep{deshpande2020afmc,deshpande2021}.
We consider the simplest model of a wall-bounded flow comprising wall-scaled eddies as the only eddying motions, which are represented by hairpin-type vortical structures.
It is important to note here, however, that the concept of active and inactive motions does not depend on the shape of the vortical structures, but is rather associated with the spatial distribution of velocity signatures generated by these vortical structures, as depicted in figure \ref{fig3}(a) (estimated by simple Biot$-$Savart calculations).
The inviscid nature of the wall-scaled eddy model permits slip at the wall (\emph{i.e.}, finite $u$ and $v$), while $w$ = 0 is enforced at $z$ = 0 to remain consistent with wall impermeability.
These differences in the boundary conditions result in} spatially $localized$ $w$-signatures (and consequently $uw$-signatures) at $z$ from any wall-scaled eddy of $\mathcal{H}$ $\sim$ $\mathcal{O}$($z$), while the $u$- and $v$-signatures extend across 0 $\lesssim$ $z$ $<$ $\mathcal{H}$ (\emph{i.e.}, \emph{non-local}; refer to figure \ref{fig3}a).
\rahul{It is these characteristics that distinguish the active and inactive contributions in the velocity flow field, despite both of them originating from the same wall-scaled vortical structures.}

\rahul{Differences in the active/inactive contributions to the Reynolds stresses can be understood by invoking \citeauthor{townsend1976}'s (1976) eddy-intensity function $I_{ij}$ (where $i$,$j$ = $u$, $v$ or $w$), depicted schematically in figure \ref{fig3}(b) for the four non-zero combinations of $i$,$j$ (refer equation \ref{eq1}).
Here, $I_{ij}$ is representative of the contribution from each eddy to the normal and Reynolds shear stresses as a function of $z$, and this is sketched for wall-scaled eddies of three different heights ($\mathcal{H}_{1}$-$\mathcal{H}_{3}$) in figure \ref{fig3}(b), as an example.
Effects of increasing $Re_{\tau}$ can be accounted for by including taller eddies of height ${\mathcal{H}}_4$ $>$ ${\mathcal{H}}_3$ and so on.
The wall-normal distributions of $I_{ij}$ are inspired by the individual velocity distributions in figure \ref{fig3}(a), and they are used to explain the cumulative contributions to the Reynolds stresses (from the three eddies) at three different probe locations in figure \ref{fig3}(b).
For instance, when the probe is positioned at $z$ $\sim$ $\mathcal{H}_{1}$, all three eddies will contribute to $\overline{u^2}$($z$) and $\overline{v^2}$($z$), while the contribution to $\overline{w^2}$($z$) and $\overline{uw}$($z$) would come solely from the eddy of height $\mathcal{H}_{1}$.
However, upon increasing the probe location to $z$ $\sim$ $\mathcal{H}_{2}$, contributions to the wall-parallel stresses are only made by the tallest two eddies, while those to $\overline{w^2}$($z$) and $\overline{uw}$($z$) only come from the eddy of height $\mathcal{H}_{2}$, and so on.
Connecting these observations with Townsend's definitions in $\S$\ref{aem},} the active contributions at any $z$ can be solely associated with eddies of height, $\mathcal{H}$ $\sim$ $\mathcal{O}$($z$), which will contribute to $u$, $v$, $w$ and hence $uw$, at $z$ \rahul{(indicated by red background shading in figure \ref{fig3}b)}.
On the other hand, the inactive contributions are associated with relatively tall {wall-scaled} eddies: $\mathcal{O}$($z$) $\ll$ $\mathcal{H}$ $\lesssim$ $\mathcal{O}$($\delta$). 
These contribute to $u$($z$) and $v$($z$), but not $w$($z$) and $uw$($z$) \rahul{(indicated by blue background shading)}.

\rahul{Considering that increasing $Re_{\tau}$ introduces new eddies that are much taller than $z$, only the inactive contributions to $u$($z$) and $v$($z$) will increase for all $z$ $\ll$ $\mathcal{H}$, while the active contributions will exhibit $Re_{\tau}$-invariance when scaled in wall units (\emph{i.e.}, $z$ and $U_{\tau}$).
This explains the contrasting trends exhibited by the Reynolds stresses in (\ref{eq1}), where ${\overline{u^2}}^+$($z$) and ${\overline{v^2}}^+$($z$) are $Re_{\tau}$-dependent while ${\overline{w^2}}^+$($z$) and ${\overline{uw}}^+$($z$) exhibit $Re_{\tau}$-invariance in the log-region.  
In summary, both active and inactive motions contribute to the wall-parallel velocity fluctuations, but only the former contribute to the Reynolds shear stresses and $w$-fluctuations}. This is expressed mathematically following \citep{panton2007,deshpande2021uw}:
\begin{equation}
\begin{aligned}
\label{eq2}
{u}(z) &= {u_{\rm a}}(z) + {u_{\rm ia}}(z),\\
{v}(z) &= {v_{\rm a}}(z) + {v_{\rm ia}}(z)\; \text{and} \\
{w}(z) &= {w_{\rm a}}(z),
\end{aligned}
\end{equation}
where subscripts `${\rm a}$' and `${\rm ia}$' respectively denote active and inactive components. The variances can be decomposed following \citep{panton2007,deshpande2021uw}:
\begin{equation}
\label{eq3}
\begin{aligned}
{{\overline{u^2}}^+} &= {{\overline{u^2_{\rm ia}}}^+} + {{\overline{u^2_{\rm a}}}^+} + 2\overline{{u_{\rm ia}}{u_{\rm a}}}, \\
{{\overline{v^2}}^+} &= {{\overline{v^2_{\rm ia}}}^+} + {{\overline{v^2_{\rm a}}}^+} + 2\overline{{v_{\rm ia}}{v_{\rm a}}}, \\
{{\overline{w^2}}^+} &= {{\overline{w^2_{\rm a}}}^+} \textrm{and} \\
{{\overline{uw}}^+} &= {{\overline{{u_{\rm a}}{w}}}^+} + {{\overline{{u_{\rm ia}}{w}}}^+}.
\end{aligned}
\end{equation}
While $\overline{{u_{\rm ia}}{u_{\rm a}}}$, $\overline{{v_{\rm ia}}{v_{\rm a}}}$ and $\overline{{u_{\rm ia}}{w}}$ = 0 for a traditionally conceptualized wall-scaled eddy field (\emph{i.e.}, considering only linear superposition of velocity fluctuations originating from various hierarchies of wall-scaled eddies), that is not true for an actual/real wall-bounded flow \citep{deshpande2021,deshpande2021uw}.
A real TBL also comprises very-large-scale inertial motions/superstructures that are inherently inactive per definition of Townsend \citep{deshpande2021} and interact non-linearly with/modulate the eddies \emph{local} to $z$ \citep{metzger2001,mathis2013,baars2016slse}.
The causality between superstructures and non-linear interactions, however, is a topic of ongoing debate \citep{andreolli2023}.
This, combined with the fact that these non-linear interactions are smaller in magnitude than individual $\overline{u^2_{\rm ia}}$ and $\overline{u^2_{\rm a}}$ components (\citealp{deshpande2021uw}), makes the investigation of non-linear interactions beyond the scope of this study.
\rahul{The present study aims to deploy the decomposition methodology facilitating (\ref{eq2}) on instantaneous flow fields of published data sets, to analyze for the first time the active and inactive contributions to ${\tau}_w$ and $p_w$.
This work builds on the past successes of the decomposition methodology \citep{deshpande2021} that has yielded empirical support for the scaling characteristics of the active and inactive components, hypothesized by \citet{townsend1976}.}
This includes the $z$-scaling behaviour of ${{\overline{u^2_{\rm a}}}^+}$, and the inverse logarithmic variation of ${{\overline{u^2_{\rm ia}}}^+}$ associated with the $k^{-1}$-scaling, to list a few.

\section{Data sets and methodology}
\label{data}

The present study considers two previously published multi-point data sets across a large $Re_{\tau}$ range: $\mathcal{O}$($10^3$) $\lesssim$ $Re_{\tau}$ $\lesssim$ $\mathcal{O}$($10^6$), for the SLSE analysis.
The low-$Re_{\tau}$ data set is from the high-resolution large-eddy simulation (LES) of a zero-pressure-gradient (ZPG) TBL by \citet{eitel2014}, which was computed over a numerical domain large enough for the TBL to evolve up to $Re_{\tau}$ $\approx$ 2000.
\rahul{In comparison to a fully-resolved DNS, the resolution employed in the LES data is just a factor of 2 coarser in the wall-parallel directions, and about 1.5 times coarser along the wall-normal one. 
Hence, instead of using a conventional subgrid scale (SGS) model in the simulation, a small forcing was implemented on the very small scales, essentially to add some extra dissipation. 
As a consequence, only two minor statistical differences emerge compared to a fully resolved DNS: a slight attenuation of the near-wall peak of ${\overline{u^2}}^+$, and the resolved TKE dissipation of 87.2\% of that in a conventional DNS. 
However, adding the dissipation associated with the forcing recovers 99.8\% of the TKE dissipation.
Since the present study predominantly focuses on the correlation between synchronously sampled time-series of the desired wall properties (${\tau}^+_w$, ${p}^+_w$) and the overlying flow field ($u^+$, $w^+$), the use of a high-resolution LES instead of DNS does not influence the present conclusions.}
These correlations are investigated at designated streamwise locations of the numerical domain, corresponding to $Re_{\tau}$ $\approx$ 500, 1000, 1500 and 2000.
For this, the time series of the LES were sampled across the entire domain cross-section ($y-z$), with a resolution of ${\Delta}t^+$ $\lesssim$ 0.5 and for a total eddy-turnover time (${t_{\rm samp}}{U_{\infty}}/{\delta}$) $\approx$ 243, where $U_\infty$ is the freestream speed.
Combined with the option to ensemble-average across the span, these time-series are long enough to obtain a sufficiently converged frequency spectra for capturing the inertial phenomena (demonstrated in \citealp{eitel2014}).

The high-$Re_{\tau}$ ($\sim$ $\mathcal{O}$($10^6$)) data set is from the neutrally-buoyant surface layer at the SLTEST facility in Utah \citep{marusic2007}.
The data comprises synchronously acquired time-series of ($u^+$, $w^+$) from five sonic anemometers positioned on a vertical tower, within the log region (0.0025 $<$ $z/{\delta}$ $<$ 0.0293), and wall-shear stress-signals (${\tau}^+_w$) measured using a custom-designed sensor placed vertically below the sonics.
This data was acquired at a time resolution of ${\Delta}t^+$ $\lesssim$ 78.4 and for a total of 175 eddy-turnover times, across a viscous-scaled measuring volume of 1400 (of sonics).

While the sampling intervals for both LES and SLTEST data sets are not sufficient for fully converging/resolving the very-large-scale phenomena (quantitatively), they have been analyzed previously by \citet{deshpande2024tsfp} and found to be sufficient to attain an accurate qualitative understanding.
This was demonstrated by repeating the same statistical analysis (as presented ahead in \S\ref{slse}) on the simultaneous ($u^+$,$w^+$) and ${\tau}^+_w$ measurements conducted in the large Melbourne wind tunnel \citep{deshpande2021uw}, time series for which were sampled for a much longer time interval and at greater frequency.
The results and trends derived from the wind-tunnel data were qualitatively consistent with those obtained from the LES and SLTEST data sets, which will serve as the basis for the present study.
Both these data sets provide rare access to velocity-fluctuation time series across the log-region synchronously with ${\tau}_w$, for TBLs spanning a broad $Re_{\tau}$ range ($10^3$--$10^6$).
This offers a unique opportunity to directly test the hypothesis of \citet{bradshaw1967} and \cite{giovanetti2016}, regarding TKE production and transport mechanisms from log region to the wall, and understand how they affect ${\tau}_w$ and $p_w$.
Our present conclusions will only depend on the qualitative energy variation across $z$, and not focus on its quantification/scaling (requiring convergence).
We note that although the SLTEST flows are transitionally rough TBLs, the roughness effects are insignificant beyond the roughness sublayer \citep{klewicki2008,marusic2007}.
Hence, the roughness will not influence inertial eddies, which are our primary focus.

\subsection{ {Data-driven flow-decomposition methodology}}
\label{slse}

\begin{figure}
   \captionsetup{width=1.0\linewidth}
\centering
\includegraphics[width=0.85\textwidth]{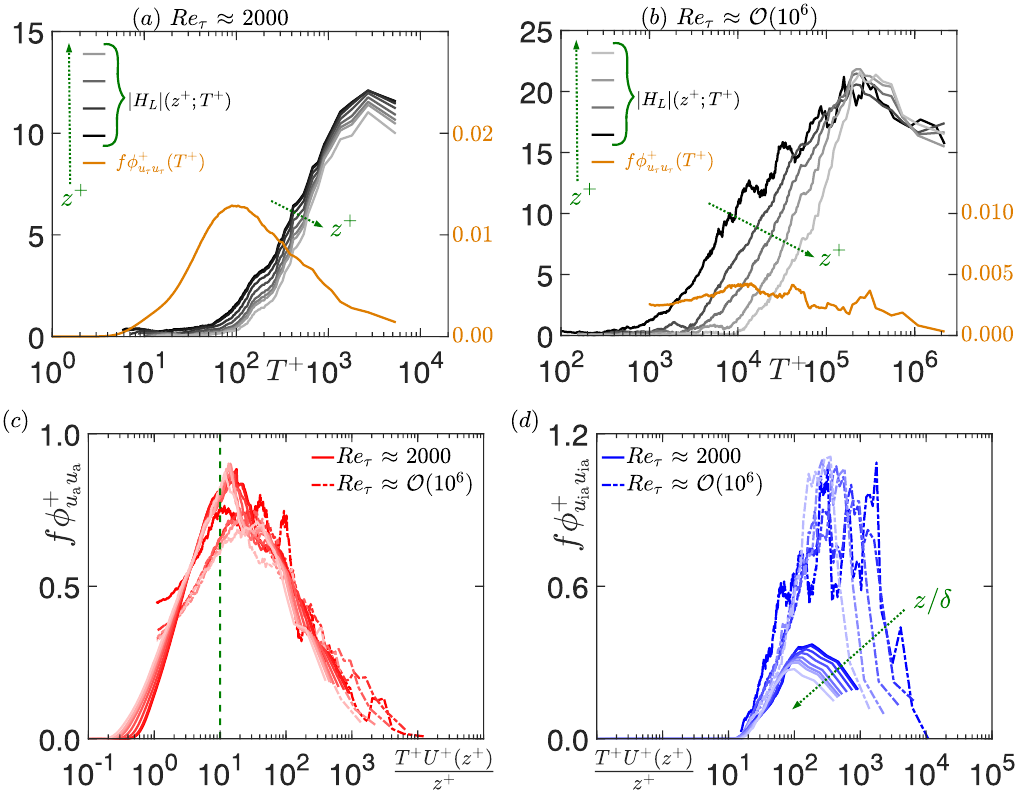}
\caption{Absolute values of the linear transfer kernel ($|H_L|$($z^+$;$T^+$); equation \ref{eq5}) computed for the (a) LES and (b) SLTEST data with grey shading representing changing $z^+$.
Here, $|H_L|$ is plotted for $z^+$ limited to the log region and its ordinate is considered on the primary vertical axis (left; in black).
While, the solid golden line represents the premultiplied spectra of the friction velocity ($f{{\phi}^+_{{u_{\tau}}{u_{\tau}}}}$) and its ordinate is considered on the secondary vertical axis (right; in golden yellow).
\rahul{(c,d) Premultiplied spectra of the active ($u_{\rm a}$) and inactive $u$-components ($u_{\rm ia}$) estimated for various $z$-locations in the log region for both the LES and SLTEST data. 
Note that the horizontal axis is chosen to test for $z$-scaling of the spectra by invoking Taylor's hypothesis.}}
\label{fig4}
\end{figure}

The multi-point nature of both LES and SLTEST data sets permit  {theoretical} estimation of $u_{\rm a}$ and $u_{\rm ia}$ by following the SLSE-based methodology proposed previously in \citet{deshpande2021} and \citet{deshpande2021uw}.
 {Throughout this paper, we limit our SLSE analysis to the log region where the concept of active and inactive motions, as well as expressions in (\ref{eq1}), have received considerable empirical support \citep{deshpande2021}.
Per the SLSE methodology} \citep{baars2016slse}, the instantaneous component $u_{\rm ia}$($z$) in the log region can be obtained by:
\begin{equation}
\label{eq4}
{\widetilde{u_{\rm ia}}}(z^{+};{T^{+}}) = {{H_{L}}(z^{+};{T^{+}})}{\widetilde{u_{\tau}}}({T^{+}}). 
\end{equation}
Here, ${\widetilde{u_{\tau}}}$($T^+$) = $\sqrt{{\widetilde{{\tau}_{w}}}/{\rho}}$ = $\mathcal{F}$($u_{\tau}$($t^+$)) is essentially the Fourier transform of the friction velocity fluctuations, $u_{\tau}$($t^+$) in time $t$, where $\rho$ is density.
${\widetilde{u_{\tau}}}$ acts as the scale-specific unconditional input required to obtain the scale-specific conditional output, ${\widetilde{u_{\rm ia}}}(z^{+};{T^{+}})$.
\rahul{The linear relationship between $\widetilde{u_{\rm ia}}$ and $\widetilde{u_{\tau}}$ is inspired by the original definitions of active and inactive motions derived from the wall-scaled eddy model (refer $\S$\ref{aem}), which assumes a linear superposition of the velocity signatures generated by individual eddies.}
Further, $H_{L}$ is the complex-valued linear transfer kernel reconstructed by cross-correlating the synchronously acquired $\widetilde{u}$($z^+$) and $\widetilde{u_{\tau}}$ following:
\begin{equation}
\label{eq5}
{H_{L}}(z^{+};{T^{+}}) = \frac{ \{ \widetilde{u}(z^{+};{T^{+}}){{\widetilde{u_{\tau}}}^{\ast}}({T^{+}}) \} }{ \{ {{\widetilde{u_{\tau}}}({T^{+}})}{{\widetilde{u_{\tau}}}^{\ast}}({T^{+}}) \} },
\end{equation}
where the curly brackets ($\{\cdot \}$) and asterisk ($\ast$) denote ensemble averaging and complex conjugate, respectively.
\rahul{Here, the definition of $H_{L}$ is underpinned by the discussion presented in $\S$\ref{aem} and figure \ref{fig3} that the inactive components coexisting at $z^+$ extend down to the wall (\emph{i.e.}, influence $u_{\tau}$), while the active components are localized to $z$.
Hence, (\ref{eq4}) and (\ref{eq5}) essentially classify $\widetilde{u_{\rm ia}}$ as a subset of the total momentum $\widetilde{u}$ that is coherent with $\widetilde{u_{\tau}}$.}
Figures \ref{fig4}(a,b) depict $|H_{L}|$(${z^+}$, $T^+$) computed from the LES and SLTEST data sets at $z^+$ within the log region, alongside the premultiplied spectra of $u_{\tau}$ (${f}{{\phi}^+_{{u_{\tau}}{u_{\tau}}}}$), where \rahul{$|{\cdot}|$} represents the modulus.
Before plotting, $|H_{L}|$ has been smoothed based on a 25\% bandwidth moving filter (BMF; \citealp{baars2016slse}).
This is done to remove noise emerging from the mathematical operations in (\ref{eq5}), which are conducted on a per-scale basis.
\rahul{It is evident that $|H_L|$ is non-zero across a broadband range of $T^+$ when $z$ is close to the lower bound of the log region.
However, $|H_L|$ gets restricted to relatively large scales with increase in $z^+$, essentially acting as a $z$-dependent low-pass filter conforming to larger scales with increasing $z$, in a manner consistent with Townsend's wall-scaled eddy hypothesis (refer $\S$\ref{aem}).
These trends also align with} $|H_L|$ computed previously using long time-series signals from the Melbourne wind-tunnel data \citep{deshpande2021uw,deshpande2024tsfp}, 
confirming they are physical and not artefacts of insufficient convergence.

Significantly, for both the LES and SLTEST data, $|H_{L}|$  extends to sufficiently large $T^+$ values where the corresponding $f{{\phi}^+_{{u_{\tau}}{u_{\tau}}}}$ is negligible beyond the considered range. This ensures proper estimation of the very-large-scale ${\widetilde{u_{\rm ia}}}$ signal.
Similarly, the time resolution and frequency response of the wall-shear-stress sensor at high $Re_{\tau}$, despite limiting the measurement of wall-shear-stress spectrum to $T^+$ $\gtrsim$ 1000 (figures \ref{fig1}a, \ref{fig4}b), is sufficient to resolve ${\widetilde{u_{\rm ia}}}$ considering $|H_{L}|$ $\sim$ 0 at $T^+$ $\lesssim$ 1000.
\rahul{The efficacy of extracting the inactive contributions is demonstrated by examining the scaling characteristics of the premultiplied $u_{\rm ia}$-spectra ($f{{\phi}^+_{{u_{\rm ia}}{u_{\rm ia}}}}$; also see \citealp{deshpande2021}).
Figure \ref{fig4}(d) depicts $f{{\phi}^+_{{u_{\rm ia}}{u_{\rm ia}}}}$ estimated for various $z/{\delta}$ in the log region for the two data sets, and it is plotted against $z$-scaled streamwise wavelengths (${T^+}{U^+}$($z^+$)/$z^+$) computed based on Taylor's hypothesis (well accepted for inertial scales below 6$\delta$; \citealp{dennis2008}).
Considering that inactive contributions at any $z$ are associated with wall-scaled eddies across $\mathcal{O}$($z$) $\lesssim$ $\mathcal{H}$ $\lesssim$ $\mathcal{O}$($\delta$), $f{{\phi}^+_{{u_{\rm ia}}{u_{\rm ia}}}}$ is expected to exhibit $z$-scaling only in the relatively small-scale range, consistent with the trends in figure \ref{fig4}(d).
Further, it is evident that the relatively large-scale contributions to $f{{\phi}^+_{{u_{\rm ia}}{u_{\rm ia}}}}$ decrease with $z/{\delta}$ across both data sets, which is also consistent with the inverse-logarithmic variation of $\overline{u^2_{\rm ia}}$ expected based on the wall-scaled eddy hypothesis (refer $\S$\ref{aem} and \citealp{deshpande2021}).}

Once ${\widetilde{u_{\rm ia}}}$ is obtained via (\ref{eq4}), its time-domain equivalent can be calculated simply by taking the inverse Fourier transform \citep{deshpande2021uw}:
${{{u}_{\rm ia}}(z^{+}; {t^{+}})}$ = ${\mathcal{F}^{-1}}({{\widetilde{{u}_{\rm ia}}}(z^{+};{T^{+}})})$.
Considering the discussion on the past hypotheses in $\S$\ref{aem}, the novel analysis here is not to establish the correlation of $u_{\rm ia}$ with ${\tau}_w$ (which is imposed by definition in equation \ref{eq4}), but to investigate the variation of simultaneously acquired $u_{\rm ia}$-signals across various $z$-locations in the log region (see $\S$\ref{drag}).
Estimation of $u_{\rm ia}$ also permits calculation of the $u_{\rm a}$-time series by simple subtraction, ${{{u}_{\rm a}}(z^{+};{t^{+}})}$ = ${{{u}}(z^{+};{t^{+}})}$ -- ${{{u}_{\rm ia}}(z^{+};{t^{+}})}$, thereby associating $u_{\rm a}$ with the $u-$subset that is incoherent with the wall (\emph{i.e.}, $u_{\tau}$).
\rahul{Similar to the investigation of $f{{\phi}^+_{{u_{\rm ia}}{u_{\rm ia}}}}$, we can examine the premultiplied $u_{\rm a}$-spectra ($f{{\phi}^+_{{u_{\rm a}}{u_{\rm a}}}}$) for its expected scaling arguments.
Per \citeauthor{townsend1976}'s (1976) hypothesis, active contributions at $z$ are solely associated with wall-scaled eddies of $\mathcal{H}$ $\sim$ $\mathcal{O}$($z$), suggesting $f{{\phi}^+_{{u_{\rm a}}{u_{\rm a}}}}$ should exhibit $z$-scaling irrespective of the flow $Re_{\tau}$.
Figure \ref{fig4}(c) depicts reasonably good $z$-scaling of the premultiplied $u_{\rm a}$-spectra across $\mathcal{O}$($10^3$) $\lesssim$ $Re_{\tau}$ $\lesssim$ $\mathcal{O}$($10^6$), consistent with the previous findings in \citet{deshpande2021}.
This ability to estimate $u_{\rm a}$ enables} computation of the Reynolds shear stresses associated exclusively with the active motions (${u_{\rm a}}w$($z^+$; $t^+$)), which should correspond closely with the net Reynolds shear stresses ($uw$($z^+$; $t^+$)) per \citeauthor{townsend1961}'s (1961) hypothesis.
These hypotheses are tested in this study using simultaneously acquired time series signals, across the log region, for the first time (see $\S$\ref{drag}).

\section{Results and discussions}
\label{results}

\subsection{Active and inactive  {contributions to} the wall-shear stress}
\label{drag}

\begin{landscape}
\begin{figure}
   \captionsetup{width=1.0\linewidth}
\centering
\includegraphics[width=1.65\textwidth]{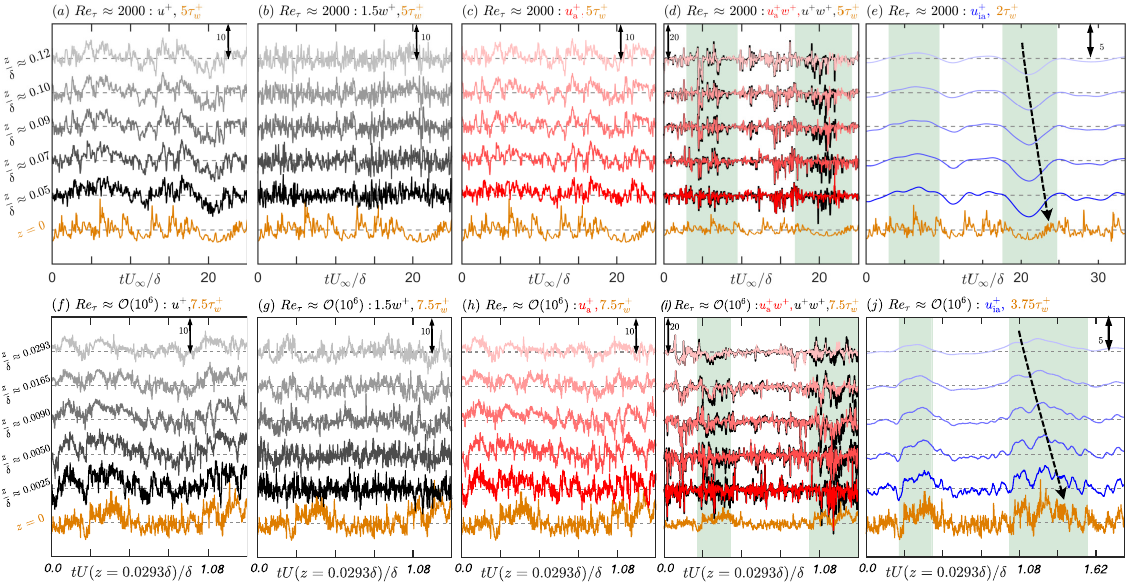}
\caption{Instantaneous (a,f) streamwise and (b,g) amplified wall-normal velocity fluctuations (in black) at various $z^+$ decomposed into their corresponding (c,h) active ($u^+_{\rm a}$; in red) and (e,j) inactive ($u^+_{\rm ia}$; in blue) components via the SLSE-based methodology discussed in $\S$\ref{slse}.
Note the different vertical offsets for various time series as well as different ordinate ranges.
(d,i) compares the time series of the full momentum flux (${u^+}{w^+}$; in black) and that associated with the active motions (${u^+_{\rm a}}{w^+}$; in red). 
Dark to light shading indicates increase in $z/{\delta}$ across all plots.
The synchronously acquired time series of the wall-shear-stress fluctuations (${\tau}^{+}_{w}$; in golden) have been plotted for: (a-e) $Re_{\tau}$ $\approx$ 2000 (LES) and (f-j) $Re_{\tau}$ $\approx$ $\mathcal{O}$($10^6$) (SLTEST) data sets and has been intentionally amplified for clarity.
Background green shading in (d,e,i,j) highlights portions of time series associated with relatively high magnitudes of $u^+_{\rm ia}$, which increase with decreasing $z/{\delta}$ indicating downward momentum transfer.}
\label{fig5}
\end{figure}
\end{landscape}

This section focuses on responding to research question (i) raised in $\S$\ref{intro}, regarding the role played by inactive motions in transporting large-scale energy from the outer region to the wall (to explain ${\tau}_w$-signatures at large $T^+$).
Figures \ref{fig5}(a,f) plot a small subset of the full $u^+$-time series sampled in the log region, synchronously with that of ${\tau}^+_w$, from the LES and SLTEST data sets respectively.
This data is used to obtain the corresponding active (figures \ref{fig5}c,h) and inactive components (figures \ref{fig5}e,j) of $u^+$ by following the procedure described in $\S$\ref{slse}.
As expected based on equation (\ref{eq4}), $u_{\rm ia}$-signals correspond predominantly to low-frequency features (\emph{i.e.}, large $T^+$) that are highly correlated with ${\tau}_w$, while $u_{\rm a}$-signals are representative of the intermediate-frequency phenomena that are uncorrelated with ${\tau}_w$.
Also plotted in figures \ref{fig5}(b,g) are the $w^+$-signals corresponding to the same instants as those considered for the $u^+$-signals.
Notably, the $w$-signals exhibit the same characteristics as those of $u_{\rm a}$ across the log region.
This is consistent with Townsend's description of $u_{\rm a}$ and $w$ being associated with {wall-scaled} eddies local to $z$ (\emph{i.e.}, $\mathcal{H}$ $\sim$ $z$), while $u_{\rm ia}$ corresponds to
\noindent relatively taller and larger  {wall-scaled} eddies ($\mathcal{O}$($z$) $\ll$ $\mathcal{H}$ $\lesssim$ $\mathcal{O}$($\delta$)) whose contributions extend to the wall (refer \ref{aem}).

\begin{figure}
   \captionsetup{width=1.0\linewidth}
\centering
\includegraphics[width=1.0\textwidth]{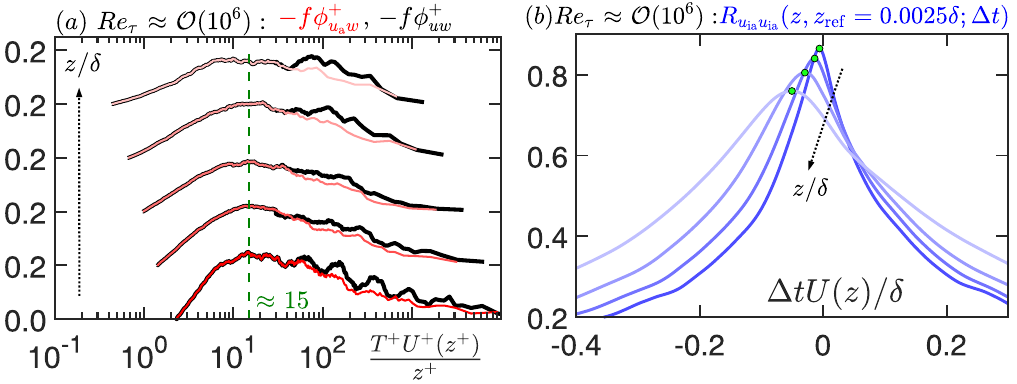}
\caption{\rahul{(a) Premultiplied co-spectra of the full Reynolds shear stress (${f}{{\phi}^+_{uw}}$; in black) compared against that associated with the active component (${f}{{\phi}^+_{{u_{\rm a}}w}}$; in red) at various $z/{\delta}$. 
Spectra profiles for increasing $z/{\delta}$ are vertically offset by 0.2 for convenience in comparison.
Dashed green line indicates $z$-scaling of the co-spectra peak at ${T^+}{U^+}(z^+)$ $\approx$ 15$z$.
(b) Cross-correlation between $u_{\rm ia}$ at $z_{\rm ref}$ = 0.0025$\delta$ and $z$ $>$ $z_{\rm ref}$, with the maximum $R_{{u_{\rm ia}}{u_{\rm ia}}}$ highlighted by green circles.}} 
\label{fig6}
\end{figure}

\citeauthor{townsend1976}'s (1976) hypothesis is mainly centred on the fact that the active components are solely responsible for the shear stresses and TKE production across log region $-$ this is clearly evident from figures \ref{fig5}(d,i) comparing ${u_{\rm a}}w$ (in red) with $uw$ (in thicker black lines).
The good overlap of the two signals, across $10^3$ $\lesssim$ $Re_{\tau}$ $\lesssim$ $10^6$, is a testament to the capability of the SLSE-based methodology of extracting the active components from simultaneously acquired signals across the log region.
\rahul{The observations from figure \ref{fig5}(i) have been reaffirmed statistically in figure \ref{fig6}(a), where we compare the premultiplied co-spectra of the full Reynolds shear stress (${f}{{\phi}^+_{uw}}$) with that associated solely with the active component (${f}{{\phi}^+_{{u_{\rm a}}w}}$), across various $z/{\delta}$ for the $Re_{\tau}$ $\sim$ $\mathcal{O}$($10^6$) data set.
The horizontal axis in this figure has been chosen to test for $z$-scaling of the co-spectra, by invoking the Taylor's hypothesis \citep{dennis2008} similar to figures \ref{fig4}(c,d).
In figure \ref{fig6}(a), ${f}{{\phi}^+_{uw}}$ and ${f}{{\phi}^+_{{u_{\rm a}}w}}$ agree very well in the intermediate-scale range for all $z/{\delta}$, with the co-spectra peak exhibiting wall-scaling at ${T^+}{U^+}(z^+)$ $\approx$ 15$z^+$.
This reinforces the association of $u_{\rm a}$ with the active components per the definition of \citet{townsend1976}.}

\rahul{There are, however, slight disagreements between ${f}{{\phi}^+_{uw}}$ and ${f}{{\phi}^+_{{u_{\rm a}}w}}$ at the larger scales. 
This can be associated with the differences observed between ${u_{\rm a}}w$- and $uw$-time series in figure \ref{fig5}(d,i)}, at instants indicated by a green background.
These are representative of the non-linear interactions between the active motions and inactive superstructures (\emph{i.e.}, ${u_{\rm ia}}w$), which become non-negligible only at time instants corresponding to high-amplitude signatures in $u_{\rm ia}$ (also highlighted by the green background in figures \ref{fig5}e,j).
\rahul{Hence, figure \ref{fig6}(a) reaffirms that the non-linear interactions between the active and inactive components contribute towards a minor subset of the Reynolds shear stresses (and consequently, the $\overline{u^2}$ production), associated with the very large scales.
These contributions, however, cannot be explained based on Townsend's wall-scaled eddy framework and is a shortcoming, requiring consideration in future upgrades of the model/framework.}

\rahul{Another notable observation from figures \ref{fig5}(e,j) is the discernible increase in $u_{\rm ia}$-fluctuation magnitudes with decreasing $z/{\delta}$}, which can be compared based on the dashed grey lines representing zero magnitude (for signals at each $z$ location). 
This suggests a wall-ward transport of streamwise momentum by the inactive components (indicated by the dashed black arrows) that is locally produced at each $z$ by the active component (conceptualized in figures \ref{fig2}c-e).
Since the correlation between $u_{\rm ia}$ and ${\tau}_w$ has already been quantified by $|H_{L}|$ via (\ref{eq5}), we know that this transport occurs all the way down to the wall.
The novel result here is the time-synchronized increment in $u_{\rm ia}$ magnitude with decreasing $z$, which is revealed only after the flow decomposition.
It presents direct evidence of the non-local transfer of momentum/energy in the wall-normal direction across a broad $Re_{\tau}$-range, which was previously noted only via spectral distributions at low $Re_{\tau}$ \citep{cho2018,mklee2019}.
\rahul{The average time delay/offset for the increase in $u_{\rm ia}$ magnitude, with decreasing $z$, can be estimated by computing the two-point correlation ($R_{{u_{\rm ia}}{u_{\rm ia}}}$) defined as:
\begin{equation}
\label{eq6}
R_{{u_{\rm ia}}{u_{\rm ia}}}({z},{z_{\rm ref}};{\Delta}t) = 
\frac{ \overline{ {{u_{\rm ia}}({z_{\rm ref}} \: {\approx} \: 0.0025{\delta}, \: {t})}\:{{u_{\rm ia}}({z}, {t} + {\Delta}t)} } }
{ \sqrt{ \overline{{u^2_{\rm ia}}({z_{\rm ref}} \: {\approx} \: 0.0025{\delta}) }} \sqrt{ \overline{{u^2_{\rm ia}}({z}) }} }.
\end{equation}
Figure \ref{fig6}(b) depicts $R_{{u_{\rm ia}}{u_{\rm ia}}}$ for the $Re_{\tau}$ $\sim$ $\mathcal{O}$($10^6$) data at various $z$ $>$ $z_{\rm ref}$.
It is evident that the maxima of $R_{{u_{\rm ia}}{u_{\rm ia}}}$, which are highlighted by a green circle, are found for a non-zero $|{\Delta}t|$ that increases with increasing $z/{\delta}$.
This can be interpreted as the average increase in time needed for the wall-normal energy transport, from $z$ to $z_{\rm ref}$.}

The significance of these empirical evidences are highlighted by quoting text directly from \citet{giovanetti2016}, who hypothesized that:
``\emph{the energy-containing motions, which essentially reside in the logarithmic and outer regions, transport the streamwise momentum to the near-wall region through
their inactive part, while generating Reynolds shear stress with their wall-detached
wall-normal velocity component in the region much further from the wall}''.
Here, \citet{giovanetti2016} use `wall-detached' to essentially refer to the active component responsible for the $w$-fluctuations, given that it does not physically extend down to the wall.
Hence, figures \ref{fig5}(d,e,i,j) cumulatively provide strong empirical evidence in support of the hypothesis proposed by \citet{bradshaw1967} and \citet{giovanetti2016}.
\rahul{They also align with the recent conjecture by \citet{mklee2024}, regarding `origin' of the large-scale ${\tau_w}-$signatures based on outer energy transported to the wall}.

Unfortunately, the data considered in this study does not permit a direct comparison of $\overline{u^2_{\rm ia}}$ with $\overline{{\tau}^2_w}$ for increasing $Re_{\tau}$, owing to mismatched $z^+$ and $z/{\delta}$ for the two data sets, alongside the under-resolved measurement of $\tau_w$ (both spatially and temporally).
Interested readers may refer to discussions and analyses presented in \citet{deshpande2021}, wherein it is suggested that the $\overline{u^2_{\rm ia}}$ at a fixed $z/{\delta}$ would be expected to increase with $Re_{\tau}$ due to the broadening of the wall-scaled eddy hierarchy (see also figures \ref{fig2}b,e and \ref{fig3}). 
This insight potentially explains the Reynolds-number dependency of the low-frequency/large-scale signatures in the ${\tau}_w$-spectra ($T^+$ $>$ 1000), as noted in figure \ref{fig1}(a) and discussed in the literature \citep{orlu2011,mathis2013}.
\rahul{These large-scale contributions, however, only correspond to a subset of the total inertial velocity signatures, \emph{i.e.} the ones that are physically coherent with the wall.
This, combined with the statistically dominant contribution from the near-wall streaks ($T^+$ $\lesssim$ $\mathcal{O}$($10^3$)), is likely responsible for the weak $Re_{\tau}$-growth of the ${\tau}_w$-spectra in the intermediate-scales ($\mathcal{O}$($10^2$) $\lesssim$ $T^+$ $\lesssim$ $\mathcal{O}$($10^3$)).}

Besides testing past hypotheses, analysis in this section also demonstrates the unique capability to segregate instantaneous flow components associated with two key energy-transfer mechanisms in high $Re_{\tau}$ wall flows $-$ TKE production and its wall-normal transport.
This is a promising development for design of real-time control strategies targeting either of these mechanisms.
However, more work is required with respect to identifying 3-D flow features/motions responsible for these mechanisms before such strategies can be actualized.
As is evident from figures \ref{fig5}(b,g), the wall-normal transport cannot be explained purely based on the $w$-time series sampled across limited (and large) wall-normal offsets. 
But it could be possible by analyzing velocity fluctuations/gradients acquired with good wall-normal resolution. 
This, however, is beyond the scope of the present study.

\subsection{Active and inactive  {contributions to} the wall-pressure}
\label{pressure}

\begin{figure}
   \captionsetup{width=1.0\linewidth}
\centering
\includegraphics[width=1.0\textwidth]{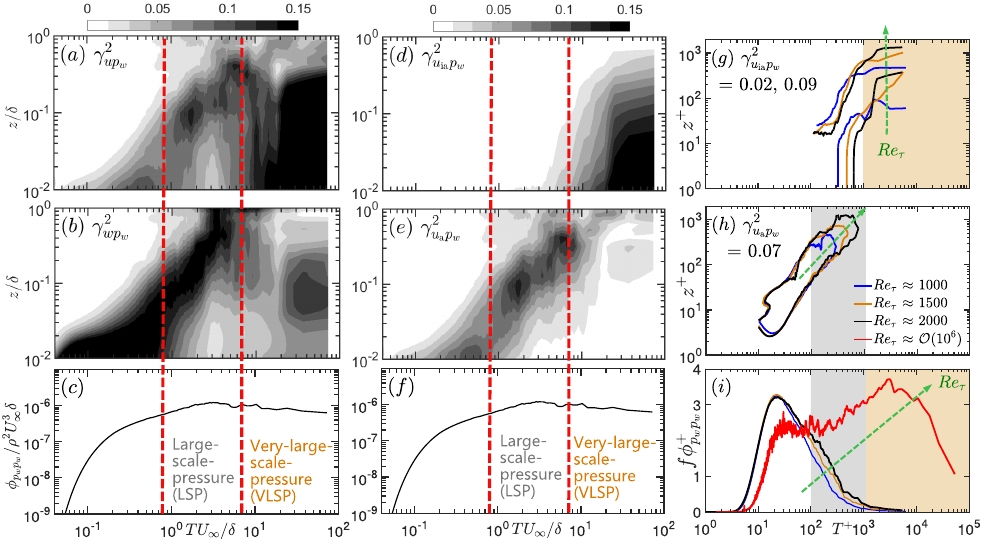}
\caption{Linear coherence spectrum (${\gamma}^2_{{u_i}{p_w}}$; equation \ref{eq6}) computed between $u_i$($z^+$) and $p_w$ as a function of outer- or inner-scaled $T$: (a) ${\gamma}^2_{{u}{p_w}}$, (b) ${\gamma}^2_{{w}{p_w}}$, (d,g) ${\gamma}^2_{{u_{\rm ia}}{p_w}}$ and (e,h) ${\gamma}^2_{{u_{\rm a}}{p_w}}$. 
(a,b,d,e) depict ${\gamma}^2_{{u_i}{p_w}}$ computed from $Re_{\tau}$ $\approx$ 2000 LES data and compared against the (c,f) outer-scaled $p_w$-spectrum plotted versus outer-scaled $T$.
(g,h) respectively depict ${\gamma}^2_{{u_{\rm ia}}{p_w}}$ and ${\gamma}^2_{{u_{\rm a}}{p_w}}$ contours for data across various $Re_{\tau}$, which are compared against (i) $f{{\phi}^+_{{p_w}{p_w}}}$ plotted versus $T^+$.} 
\label{fig7}
\end{figure}

This section focuses on addressing the research question (ii) raised in $\S$\ref{intro}, regarding the inertial motions responsible for the $Re_{\tau}$-growth of the $p_w$-spectra.
For this, we compute the linear coherence spectrum \citep{gibeau2021,baars2023} between the fluctuating wall-pressure and velocity fluctuations following:
\begin{equation}
\label{eq6}
{{\gamma}^2_{{u_i}{p_w}}}(z^{+};{T^{+}}) = \frac{| \{ \widetilde{u_i}(z^{+};{T^{+}}){{\widetilde{p_{w}}}^{\ast}}({T^{+}}) \} |^2}{{ \{ {{\widetilde{u_i}}({z^{+};T^{+}})}{{\widetilde{u_i}^{\ast}}(z^{+};{T^{+}}) \} }}{ \{ {{\widetilde{p_w}}(T^{+})}{{\widetilde{p_w}^{\ast}}({T^{+}}) \} }}},
\end{equation}
where $u_i$ can be either $u$, $w$, $u_{\rm ia}$ or $u_{\rm a}$, (|$\cdot$|) represents absolute values, while all other notations are the same as in equation (\ref{eq5}).
Note that ${\gamma^2_{{u_i}{p_w}}}$ can be interpreted as the spectral equivalent of a physical two-point correlation between $p_w$ and $u_i$, which varies between 0 $\le$ ${\gamma^2_{{u_i}{p_w}}}$ $\le$ 1 by definition.
Furthermore, ${\gamma^2_{{u}{p_w}}}$ and ${\gamma^2_{{w}{p_w}}}$ have been analyzed previously for a canonical TBL at $Re_{\tau}$ $\approx$ 2000 by \citet{gibeau2021}, to establish the scale-based coupling between the inertial motions and the large-scale pressure (LSP; 0.8 $<$ $T{U_{\infty}}/{\delta}$ $<$ 7), as well as the very-large-scale pressure (VLSP; $T{U_{\infty}}/{\delta}$ $>$ 7) regions of the $p_w$-spectrum.
To validate our analysis, we recompute ${\gamma^2_{{u}{p_w}}}$ and ${\gamma^2_{{w}{p_w}}}$ (figures \ref{fig7}a,b) from the LES data set at $Re_{\tau}$ $\approx$ 2000 and compare it with the outer-scaled $p_w$-spectrum (figure \ref{fig7}c).
The coherence spectra depicts characteristics consistent with those reported by \citet{gibeau2021}: the VLSP region is associated with relatively strong coherence between $u$ and $p_w$, but not between $w$ and $p_w$. While in case of LSP, reasonable coherence can be noted between both $u$-$p_w$ and $w$-$p_w$.
If the same spectrograms were plotted as a function of streamwise wavelengths based on Taylor's hypothesis (\emph{i.e.}, ${\lambda}^+_x$ $=$ ${T^+}{{{U}}^+}$($z^+$)), the energetic ridges of ${\gamma^2_{{u}{p_w}}}$ and ${\gamma^2_{{w}{p_w}}}$ would respectively exhibit distance-from-the-wall scalings: ${{\lambda}_x}/z$ = 14 and ${{\lambda}_x}/z$ = 8.5 (not shown here), consistent with \citet{baars2023}.
Exhibition of this scaling confirms the association of the $p_w$-spectrum with the  {wall-scaled} eddy hierarchy coexisting in the inertial/large-scale range.

Connecting the differences between ${{\gamma}^2_{{u}{p_w}}}$ and ${{\gamma}^2_{{w}{p_w}}}$ with the characteristics of active and inactive components discussed in $\S$\ref{aem}, it can be hypothesized that the $p_w$-spectrum in the LSP and VLSP ranges, is respectively, correlated to the active and inactive motions.
The availability of decomposed $u_{\rm a}$ and $u_{\rm ia}$ components, for the same data as in figures \ref{fig7}(a,b), permits us to directly test this hypothesis by analyzing the ${{\gamma}^2_{{u_{\rm ia}}{p_w}}}$ and ${{\gamma}^2_{{u_{\rm a}}{p_w}}}$ results of figures \ref{fig7}(d,e), respectively.
It is evident that the coherence between $u_{\rm ia}$-$p_w$ is limited to the VLSP range, while that between $u_{\rm a}$-$p_w$ (in the inertial region) is predominantly in the LSP range, similar to ${{\gamma}^2_{{w}{p_w}}}$. 
All the observations are consistent with our hypothesis above, with the similarity between ${{\gamma}^2_{{w}{p_w}}}$ and ${{\gamma}^2_{{u_{\rm a}}{p_w}}}$ expected considering $w$-signatures are predominantly active contributions (see equation \ref{eq2}). 
The present results offer phenomenological explanations for the conclusions drawn by  \citet{gibeau2021}, who associated $p_w$-fluctuations in the LSP range with Reynolds shear stress-carrying ejection and sweep events in the log region, \emph{i.e.}, the active components of the wall-scaled eddies. 
Conversely, $p_w$-fluctuations in the VLSP range were linked with the large $\delta$-scaled motions and superstructures, which correspond to the inactive components \rahul{transporting energy from outer region to the wall.
The results also confirm the dependency of $p_w$ on velocity signatures that are coherent (like, inactive) as well as incoherent (like active) with the wall, consistent with the pressure Poisson equation \citep{tsuji2007}}.

Since the wall-scaled eddy hierarchy grows by coexisting in a physically taller log region with increasing $Re_{\tau}$ (figures \ref{fig2}b,e; \citealp{deshpande2021}), the association of the LSP and VLSP ranges, with $u_{\rm a}$ and $u_{\rm ia}$, can be used to explain the $Re_{\tau}$-variation of the $p_w$-spectrum.
To this end, figures \ref{fig7}(g,h) respectively plot the constant energy contours of ${{\gamma}^2_{{u_{\rm ia}}{p_w}}}$ and ${{\gamma}^2_{{u_{\rm a}}{p_w}}}$ versus $T^+$ for 1000 $\lesssim$ $Re_{\tau}$ $\lesssim$ 2000, each of which exhibit their unique $Re_{\tau}$-trends.
Notably, the contours of ${{\gamma}^2_{{u_{\rm a}}{p_w}}}$ can be seen to grow and widen across a larger $T^+$ range with $Re_{\tau}$ (indicated by green arrows), which explains the $Re_{\tau}$-growth of the $p_w$-spectrum in the intermediate-scale range: $\mathcal{O}$($10^2$) $\lesssim$ $T^+$ $\lesssim$ $\mathcal{O}$($10^3$).
It suggests that a physically taller log region comprises a broader hierarchy of \rahul{TKE-producing} active motions, which subsequently increases contributions to the $p_w$-spectrum.
This is analogous to the broadening of the $\overline{uw}$-plateau ($\sim$ --1) in the log region with increasing $Re_{\tau}$ (equation \ref{eq1}, \citealp{marusic1995,baidya2017}).
In contrast, the active motions do not contribute to the ${\tau}_w$-spectrum since they do not physically extend down to the wall, and consequently do not influence the near-wall velocity gradient \citep{giovanetti2016}.
\rahul{It is thus obvious that the $Re_{\tau}$ trend of $f{{\phi}^{+}_{{{p}_w}{{p}_w}}}$ is governed by a broader range of inertial motions than those influencing $f{{\phi}^{+}_{{{\tau}_w}{{\tau}_w}}}$,
likely explaining the stronger $Re_{\tau}$-growth of the former compared to the latter.
Note that arriving at this realization was made possible by decomposing the $u$-$p_w$ coherence into its active and inactive components.}

Similar to the behaviour of ${{\gamma}^2_{{u_{\rm a}}{p_w}}}$, the contours of ${{\gamma}^2_{{u_{\rm ia}}{p_w}}}$ can also be seen extending across a larger $z^+$-range with $Re_{\tau}$.
This is predominantly associated with the energization and increasing wall-normal extent of the large wall-scaled eddies and superstructures \citep{mathis2013,mklee2019}.
At very high $Re_{\tau}$ $\sim$ $\mathcal{O}$($10^6$), these energetic superstructures along with a broad hierarchy of wall-scaled eddies (figure \ref{fig2}e) contribute to the significantly enhanced VLSP signatures of the $p_w$-spectrum through their inactive component (figure \ref{fig7}i).
The present analysis, hence, encourages future detailed multi-point measurements and high-fidelity simulations that can quantify the pressure-velocity coupling at high-$Re_{\tau}$ ($\gtrsim$ $10^4$), with particular focus on velocity signatures superimposed by the inertia-dominated motions.
\rahul{Consequently, it also motivates high-$Re_{\tau}$ investigations of the non-linear interactions between the large-scale velocity signatures and $p_w$ \citep{thomas1983,tsuji2015}, which have previously been limited to $Re_{\tau}$ $\lesssim$ $\mathcal{O}$($10^3$).}

\section{Concluding remarks}
\label{conclude}

{This study clarifies the $Re_{\tau}$-dependent, inermediate and large-scale contributions to the ${\tau}_w$- and $p_w$-spectra by invoking the attached- (wall-scaled) eddy framework \citep{townsend1976}, based on active and inactive motions.}
Unique multi-point data sets are analyzed using an energy decomposition methodology to reveal the contributions from TKE-producing (active) and non-producing (inactive) components, at any wall-normal position in the log-region, to the ${\tau}_w$- and $p_w$-spectra.
Inactive components of wall-scaled eddies are found to be responsible for the non-local wall-normal transport of the large-scale inertia-dominated energy, from its origin in the log region (\emph{i.e}, produced by active components), to the wall. 
\rahul{On the other hand, the majority of the TKE-producing (\emph{i.e.}, active) motions in the log region are localized and physically detached from the wall, not influencing ${\tau}_w$.}
These results provide strong empirical evidence for the hypotheses proposed by \citet{bradshaw1967} and \citet{giovanetti2016}.
They also explain the appearance \rahul{and $Re_{\tau}$-growth} of the large-scale signatures in the ${\tau}_w$-spectra, \rahul{despite the insignificant large-scale TKE production occurring in the near-wall region.}
In terms of their contribution to wall pressure, active and inactive components are respectively correlated with the intermediate and large-scale portions of the $p_w$-spectrum. Both components exhibit growth with increasing  $Re_{\tau}$, which is attributed to the broadening of the wall-scaled eddy hierarchy. 
\rahul{This is potentially responsible for the rapid $Re_{\tau}$-growth of the $p_w$-spectra relative to ${\tau}_w$ (figure \ref{fig1}a,b).}
These phenomenological explanations hold promise regarding future real-time control strategies aimed at achieving energy-efficient drag and flow-noise reduction.

While this study demonstrates the new found capability to extract the instantaneous active and inactive components, and also shedding light on their role in the large-scale energy-transfer mechanisms, it does not delve into the non-linear interactions between these motions. 
These interactions, which have been previously discussed in the literature to a certain extent \citep{cho2018,mklee2019}, can likely explain the transfer of TKE produced by active components to the inactive component of the wall-scaled eddies, before being transported downward to the wall (figure \ref{fig2}). 
\rahul{At this stage, these interactions are not accounted within the Townsend's wall-scaled eddy framework, which is inherently limited to the linear superposition of velocity signatures induced by the eddy structures.
It does, however, highlight the opportunity for future upgrades/modifications to the model by incorporating the inter-scale interactions between the wall-scaled eddy hierarchy.}



\section*{Acknowledgments}
R. D. is supported by the University of Melbourne's Postdoctoral Fellowship and acknowledges insightful discussions with Drs. W. J. Baars, B. Gibeau and S. Ghaemi. Funding is gratefully acknowledged from the Office of Naval Research: N62909-23-1-2068 (R.D., I.M.), and to R.V. from the European Research Council grant no. `2021-CoG-101043998, DEEPCONTROL'. 
Views/opinions expressed are however those of the author(s) only and do not necessarily reflect those of the European Union or the European Research Council. 
Neither the European Union nor the granting authority can be held responsible for them.

\section*{Declaration of Interests} 

The authors report no conflict of interest.

\bibliographystyle{jfm}
\bibliography{ActiveInactiveEffects_OnWall_bib}

\end{document}